\documentclass[11pt,twoside]{amsart}

\usepackage{color}
\usepackage{graphicx}

\usepackage{amssymb}
\usepackage{latexsym}

\makeatletter

\@addtoreset{equation}{section} \makeatother
\newtheorem{theorem}{Theorem}[section]
\newtheorem{remark}[theorem]{Remark}
\newtheorem{lemma}[theorem]{Lemma}
\newtheorem{proposition}[theorem]{Proposition}
\newtheorem{corollary}[theorem]{Corollary}
\newtheorem{definition}[theorem]{Definition}
\newtheorem{example}[theorem]{Example}

\def\1{\mathbf{1}}
\def\:{\lrcorner}
\def\#{\sharp}

\def\l{\lambda}

\def\g{\gamma}

\def\o{\circ}

\def\qed{\ensuremath{\quad\Box\quad}}

\def\inv#1{\raise.1em\hbox to 0pt{$^{-1}$\hss}_{#1}\;}

\def\v{\noindent}

\newcommand{\bean}{\begin{eqnarray*}}
\newcommand{\eean}{\end{eqnarray*}}
\newcommand{\benu}{\begin{enumerate}}
\newcommand{\eenu}{\end{enumerate}}
\newcommand{\eea}{\end{eqnarray}}
\newcommand{\bea}{\begin{eqnarray}}

\newtheorem{Theorem}{Theorem}

\newtheorem{Definition}{Definition}
\newtheorem{Example}{Example}

\newtheorem{Lemma}{Lemma}

\newcommand{\be}{\begin{equation}}
\newcommand{\ee}{\end{equation}}

\newcommand{\N}{{\mathbb N}}

\newcommand{\R}{{\mathbb R}}

\newcommand{\ben}{\begin{enumerate}}
\newcommand{\een}{\end{enumerate}}
\newcommand{\bit}{\begin{itemize}}
\newcommand{\eit}{\end{itemize}}
\newcommand{\edoc}{\end{document}}

\newcommand{\bdefi}{\begin{definition}}
\newcommand{\btheo}{\begin{theorem}}
\newcommand{\bprop}{\begin{proposition}}
\newcommand{\brema}{\begin{remark}}
\newcommand{\bcoro}{\begin{corollary}}
\newcommand{\blemm}{\begin{lemma}}
\newcommand{\bexam}{\begin{example}}

\newcommand{\edefi}{\end{definition}}
\newcommand{\etheo}{\end{theorem}}
\newcommand{\eprop}{\end{proposition}}
\newcommand{\erema}{\end{remark}}
\newcommand{\ecoro}{\end{corollary}}
\newcommand{\elemm}{\end{lemma}}
\newcommand{\eexam}{\end{example}}

\begin{document}

\setcounter{page}{1}

\title{Horizons}

\author[Olaf M\"uller]{Olaf M\"uller}
\address{Fakult\"at f\"ur Mathematik, Universit\"at Regensburg, Germany}
\email{olaf.mueller@mathematik.uni-regensburg.de}

\date{\today}

\maketitle

\begin{abstract}
\v We define different notions of black holes, event horizons and Killing horizons for a general time-oriented manifold $(M,g)$ extending previous notions but without the assumption of the existence of a causal boundary for $(M,g)$. The notions of 'horizon' are always conformally invariant while the notions of 'black hole' are genuinely geometric. Some connections between the different notions are found. Finally, we state and compare different versions of the weak cosmic censorship conjecture with precise geometric assumptions. 
\end{abstract}

\v Studying the history of the weak cosmic censorship conjecture, one notices sooner or later that, even in the case of asymptotically flat spaces, some notions and definitions diverge between different researchers, and the statement to show varies considerably. For non-asymptotically-flat (but still, say, globally hyperbolic) spacetimes the situation gets even worse. It is not at all clear what the precise conjectured statement should be for this case, as one ingredient, future null infinity, is not defined. This article, by using the definition of 'causal boundary' as given in \cite{FHS} (based on many previous approaches), tries to remedy the situation a bit by providing some more or less plausible definitions of the terms 'horizon' and 'black hole' applicable in every globally hyperbolic manifold (and, of course, extending the previous ones) and by showing some connections between them.

\bigskip

\v  Let us first recall the statement of the weak cosmic censorship conjecture as it is stated usually. The conjecture assures that 'singularities should be hidden behind event horizons', in other words: For a maximally globally hyperbolic manifold, incomplete timelike geodesics should not be visible from future null infinity. However, one has to put additional physical requirements like energy conditions to prevent examples like the following from happening:

\begin{Example}
\label{wrong}
Consider $(\R^2 \setminus \{ x \in \R^2 \vert u(x), v(x) >0 \}, g_0= du dv) $ and the metric $g:= f \cdot g_0$ with $f:= 1 + \psi (u) \cdot \phi (v) + \phi (u) \cdot \psi (v) $ where $\phi \in C^{\infty} (\R, [0, \infty))$ with ${\rm supp} (\phi ) =  [0, \infty)$ and $\phi (x) = x $ for all $ x \geq 1 $ and $\psi \in C^{\infty} ((- \infty, 0), [0, \infty))$ with ${\rm supp} (\psi) = [-1,0)$ and ${\rm lim}_{x \rightarrow 0} \psi (x) = \infty$. Then $(\R^2,g) $ is globally hyperbolic, maximal as a Lorentzian manifold, and the negative $t$ axis is an incomplete geodesic contained in the past of every other inextendable future timelike curve, in particular in the past of the hyperboloids $u \cdot v = {\rm const} >0$ which have infinite length. The scalar curvature of any such $(M,g)$ (for each choice of $\phi, \psi$) changes its sign (therefore violating the energy condition).
\end{Example}

\v Usually, 'future null infinity' is defined for asymptotically flat spacetimes only. Now we are going to present some more general notions of 'future null infinity', or equivalently, of 'horizon', to be able to formulate the conjecture in non-necessarily asymptotically flat contexts as well. We will use (implicitely) the notion of 'causal boundary' as defined in \cite{FHS}. All notions of 'horizon' will be conformally invariant, so they only depend on the causal structure and not on other geometrical data. Given a time-oriented manifold, we will speak of '{\em the} black hole of $(M,g)$' or '{\em the} event horizon of $(M,g)$' as those sets are uniquely given by the manifold.

\bigskip

\v {\bf General assumption:} From now on, $(M,g)$ always denotes a time-oriented connected Lorentzian manifold.

\begin{Definition}
A {\bf CITIF (in $M$)} is a piecewise $C^1$ future directed timelike curve in $M$ which is $C^0$-inextendible to the future. We introduce an order $< $ of CITIFs by proper inclusion of their past, that is, $c<k$ if $I^- (c) \subsetneq I^- (k)$. A CITIF $c$ is called {\bf non-dominated} (resp. {\bf non-dominant}) iff there is no other CITIF $k$ with $c<k$ (resp. $c>k$). A point $p$ is called {\bf upper-shielded} (resp. {\bf lower-shielded}) iff every CITIF from $p$ is non-dominated (resp. non-dominant). Finally, the {\bf upper-shielded horizon} $u_M$ (resp. {\bf lower-shielded horizon} $ l_M$) of $M$ is the boundary of the subset $U_M$ of upper-shielded (resp. $L_M$ of lower-shielded) points in $M$.
\end{Definition}

\v {\bf Remark:} Of course the sets $L_M$ and $U_M$ are future sets. Assume that $p \in L$ and that there is a CITIF $c: [0, \infty) $ from $p$ which is dominated by another CITIF $k: [0, \infty) \rightarrow M$, then there is a $t \in \R$ such that $ c(1) \in I^- (k(t))  $, thus there is a CITIF $\gamma $ starting at $p$ with $ \gamma \vert_{[0, \infty)} = k  \vert_{[0, \infty)} $ which is dominating, contradicting the assumption that $p \in L$. Therefore no CITIF from $p$ is dominated, and  $L_M \subset U_M$. 

\bigskip

\v Recently, the definition of the causal boundary of stably causal space-times has been systematized in a comprehensive way in \cite{FHS}. The causal boundary is a powerful tool e.g. for recognizing the structure of conformal embeddings of one space-time to another. However, for general time-oriented Lorentzian manifolds it does not exist. Now, we want to compare the notion of upper-shielded horizons to the usual definition of 'event horizon' in terms of the conformal boundary which is in some sense a special case of the causal boundary. Therefore, let us revise the definition of the latter.  First we consider the set of all indecomposable past sets ({\bf indecomposable} means here that the set is not the proper union of two past sets), short {\bf IP}s. By a theorem due to Geroch, Kronheimer and Penrose, every IP is either of the form $I^-(p)$ in which case it is called {\bf proper IP} or {\bf PIP} for short, or else it is of the form $I^- (c)$ for $c$ an inextendible future timelike curve $c$ in which case it is called a {\bf TIP}. The set of all IPs is called $ \hat{M} $, correspondingly the set of all IFs is called $\check{M}$. Let the common future $C^+$ resp. the common past $C^-$ of a set $X$ denote the set of points $ C^{\pm} (X) := \{ p \in M \vert p \in I^{\pm} (x) \ \forall x \in X \}  $. Now recall that $\check{M} $ and $\hat{M}$ are subsets of the potence set of $M$ and set $\check{M}_1 := \check{M} \cup \emptyset $ as well as $\hat{M}_1 := \hat{M} \cup \emptyset $. Then set $\tilde{M} := \check{M}_1 \times \hat{M}_1 \setminus (\emptyset, \emptyset)$. On $\tilde{M}$ we consider an equivalence relation $\sim$ defined by $P \sim F$ if and only $F \subset C^+ (P)$, $ F  $ maximal element w.r.t. inclusion under all indecomposable future sets in $C^+ (P)$, and $P \subset C^- (F)$, $ P  $ maximal element w.r.t. inclusion under all indecomposable past sets in $C^- (F)$. Finally we define $ \overline{M} := \{ (P, F) \in \hat{M}_1 \times \check{M }_1  \vert P \sim F \}$. Obviously $M \subset \overline{M}$, as $I^- (x) \sim I^+ (x)$, for all $x \in M$. On this point set $\overline{M}$ one defines a chronology relation which extends the one on $M$: ${\rm chr} ((P,F), (P',F')) \Leftrightarrow  F \cap P' \neq \emptyset $. Now define the {\bf naive future null infinity} $\mathcal{J}_n^+$ as the set of non-maximal elements w.r.t. ${\rm chr}$ in the image of the quotient of the future boundary, where a point $p \in \overline{M}$ is called non-maximal if and only if there is $q \in \overline{M}´\setminus \{ p \}$ s.t. for every $x $ with $chr(x,p)$ we have $chr(x,q)$. Clearly, this extends the notion of $\mathcal{J}^+$ as the lightlike part of the future boundary of an open conformal embedding (as used in \cite{EH} for the class of weakly asymptotically simple and empty spacetimes). Now, maximal elements are necessarily of the form $(P, \emptyset)$: Assume for an element of the form $(P, F \ni x)$ we can choose some $y \in I^+ (x)$ and then $ {\rm chr} ( (P, F), (I^- (y) , I^+ (y))  $. Therefore the set of non-maximal elements can be identified with the TIPs which are pasts of non-dominating CITIFs, and we get immediately the following theorem: 

\begin{Theorem}
Let $(M,g)$ have a causal boundary $\mathcal{J}$. Then $ U_M  =  M \setminus I^- (\mathcal{J}^+_n)$. \hfill \qed
\end{Theorem} 
  
\v So, in case that there is a causal boundary of $M$ as in \cite{FHS}, the definition of 'upper-shielded horizon' coincides with the usual one $\partial ( I^- (\mathcal{J}^+_n) )$. On the other hand, this new notion of 'horizon' is applicable in the much wider context of general time-orientable Lorentzian manifolds. However, most physicists would probably not consider our definition of shielded horizons the correct one. As it depends only on the causal structure, it cannot take into account sufficiently what future infinity is. As an example, modify the usual conformally equivalent representation of the two-dimensional Schwarzschild space-time as a hexagonal subset of two-dimensional Minkowski spacetime by adding or removing a small triangle of the surrounding Minkowski spacetime to future spacelike infinity. Then the original Schwarzschild event horizon is not a shielded horizon any more as there are CITIFS in the enclosed region which dominate other curves depending on their endpoint on the new future lightlike boundary. Still, those curves do stay in the Schwarzschild horizon and do certainly not satisfy our expectations of a curve 'escaping to infinity'. To remedy this difficulty and to distinguish a class of geodesics approaching infinity, we call a CITIF $c$  {\bf horizontal} iff there is no sequence $c_n$ of CITIFs with $c_n \nearrow c$, that is, $c_n < c$ for all $n \in \N$ and ${\rm lim}_{n \rightarrow \infty} c_n = c$ where convergence of the pasts $I^- (c_n)$ is defined as final coincidence with $I^- (c) $ on every compact subset, that is, for every $K \subset M$ compact there is an $m \in \N $ with $I^- (c_i) \cap K = I^- (c) \cap K $ for all $i >m$. Then a CITIF {\bf belongs to causal future null infinity} iff its past does not contain a horizontal CITIF. Finally, $E_M$ is defined as the complement of the past of causal future null infinity and the {\bf event horizon $e_M$ of $M$} as its boundary $e_M := \partial E_M$.

\bigskip

\v One theorem which will be important for the order relation between CITIFs is the following 'catcher theorem':

\begin{Theorem}
\label{buckelkurve}
Let $(M,g)$ have noncompact Cauchy surfaces. Then from any point $p \in M$ and any point $q \in I^+ (p)$ there is a CITIF $c$ from $p$ not intersecting $J^+(q)$. In particular, $I^- (c) \neq M$.  
\end{Theorem}

%\v {\bf Proof.} We use $\emptyset \neq \partial I^+ (q) \in I^+ (p)$ and openness of $I^+ (p) $. What we show is that for any two Cauchy surfaces $S,T$ such that $p \in S$ and $ q \in I^+ (S) \cap  I^- (T) $, one finds a curve $c$ from $p$ to $T$ disjoint from $I^+(q)$ (the rest is induction over the level surfaces of a Cauchy time function). To this purpose, observe that $\partial I ^+ (q) \cap I^- (T)$ is compact, thus we can cover it by finitely many convex neighborhoods $U_1,... U_m$. Choose an $n$-tuple $(q =:q_1, q_2,... q_n \in I^+(T))$ with $q_i \in \partial I^+ (T)$, $q_{i+1} \in J^+ (q_i)$ and a map $f$ from $\N_n$ to $\N_m$ with $ q_i \in U_{f(i)} \cap U_{f(i+1)}  $. Then we can choose $p_i \in U_{f(i)} \cap U_{f(i+1)} $ with $p_i \in I^-(q_i)$ and join them by broken geodesics. \hfill \qed

\v {\bf Proof.} Choose intermediate points $ z,r \in M $ such that $ p \ll z \ll r \ll s \ll q $. Choose a smooth Cauchy temporal function $t$ for $(M,g)$, and for any $a \in \R$ put $S_a := t^{-1} (\{ a \})$, $J_a := J^+(r) \cap S_a$ and $I_a := I^+ (z) \cap S_a$. Then, for any $a \in \R$, we have $I_a$ open in $S_a$, $J_a $ compact, and $J_a \subset I_a$ as a consequence of the first push-up lemma as presented in \cite{pC}, Lemma 2.4.14. Therefore, for any $a \in \R$ we have that the open set $I_a \setminus J_a$ is nonempty, as otherwise $I_a = J_a $ would be open and compact and therefore equal to $S_a$ which would be in contradiction to noncompactness of $S_a$. Therefore there are timelike future curves $c_n$ from $z$ to $p_n \in I_n \setminus J_n $, for $ n \in \N, n > t(z)$. The limit curve lemma assures that there is a causal curve $c$ starting at $z$ whose intersection with $S_n$ is nonempty and contained in the closure of $I_n \setminus J_n $, for every $ n \in \N, n > t(z) $. By considering $A_a := S_a \cap I^+(s) $ and $ B_a := S_a \cap J^+ (q) $ in addition similarly, it is easy to see that $B_a \subset {\rm int} (J_a) $, for any $a$, thus $c$ is disjoint from $J^+(q)$. Finally, we consider a future timelike curve $k$ from $p$ to $z$ and apply to the curve $l:= c \o k$ and to the neighborhood $I(p) \setminus J(q)$ the (time-dual of the) second push-up lemma as presented in \cite{pC}, Lemma 2.9.10, to conclude that there is a $C^0$-inextendible future timelike curve $m$ from $p$ disjoint to $J^+(q)$, as required. \hfill \qed  

\bigskip

\v Led by the consideration of the hexagonal form of Kruskal space-times versus causal diamonds $D_{pq} := I^+ (p) \cap I^-(q)$ in Minkowski space which are conformal to Minkowski space itself, one could think of defining horizons by the property of $D_{pq}$ that the future of each two points intersect. This is done in the notion of 'synopticity horizon':

\begin{Definition}
A subset $U \subset M $ is called {\bf synoptic} iff $I_M^+ (p) \cap I_M^+ (q) \cap U  \neq \emptyset$ for each two points $p,q \in U$. A {\bf synopticity region} is the complement of the closure of a {\bf maximal synoptic subset}, that is, of a subset $A \subset M$ which is synoptic and which is not a proper subset of a synoptic subset. Finally, a {\bf synopticity horizon} is the boundary of a synopticity region.
\end{Definition}

\v The definition implies directly that any synoptic subset is arcwise connected and any maximal synoptic subset of a globally hyperbolic manifold is open. The limiting case of de Sitter space-time shows two facts: Firstly, as opposed to the other notions of 'horizon' appearing in this article, maximal synoptic subsets are not unique: a spatial rotation maps one to another. Secondly, de Sitter space-time itself is not synoptic although it does not have a synopticity region due to the definition via complements of closures. On the other hand, the following theorem shows that every time-oriented Lorentzian manifold does contain a nonempty maximal synoptic region. Consequently, it does not contain a synopticity horizon if and only if the closure of such a maximal synoptic region is all of $M$. 

\begin{Theorem}
Every point $p \in M$ has a globally hyperbolic and synoptic neighborhood.
\end{Theorem}

\v {\bf Proof.} Let $U$ be a globally hyperbolic and geodesically convex neighborhood of $p$ (which is well-known to exist), then every local causal diamond $ I_U^+ (x) \cap I_U^- (y) $ containing $p$ and contained in $U$ satisfies the condition (because every $ I^- (z) $ is open and because $I^-$ is continuous as a set-valued map). \hfill \qed

\bigskip

\v The relation $<$ restricted to CITIFs in a maximal synoptic subset has a maximal element:

\begin{Theorem}
\label{champion}
Let $A$ be a maximal synoptic subset, then there is a CITIF $k$ with $I^- (k) =A$. 
\end{Theorem}

\v {\bf Proof.} Take a countable  covering of $A$ by sets of the form $I^+(y_n) \cap I^- (x_n) $ for $x_n, y_n \in A$ and choose $p_{n+1} \in I^+ (p_n ) \cap I^+ (x_n) \cap U$. Then join the $p_n$ by future timelike arcs. The result is a curve $k$ with $\overline{I^- (k)} = A$. It has to be a CITIF, as every extension of it to the future could be added to the subset preserving its synopticity, which contradicts the maximality assumption. \hfill \qed

\begin{Theorem}[Causal ladder L-U-E-S for horizons]
\label{lues}
We have the inclusions $L_M \subset U_M \subset E_M $, and for every point $p \in M \setminus L $ there is a synopticity region $ S $ with $L \subset S$ and $p \notin S$. 
\end{Theorem}

\v {\bf Proof.} We had already seen the first inclusion. For the second, let $p \in U_M$. Then, for all CITIFs $c$ from $p$, there is no CITIF $k$ with $I^- (c) \subsetneq I^- (k) $. But then $I^- (c)$ contains a horizontal CITIF, as $ c$ itself is horizontal: If $c_n$ were a sequence with $I^-(c_n) \rightarrow I^- (c) $, $I^- (c_n ) \subset I^- (c)$, then $I^- (c_m) \ni p$ for some $m$ due to the notion of convergence, so this $c_m$ would be a dominated CITIF from $p$, contradicting the assumption. For the third implication, let $c$ be a dominating CITIF from $p$. Then the image of $c$ is a synoptic subset and can therefore be extended to a maximal synoptic subset $A$. Because of Theorem \ref{champion}, we know that $A = I^- (k)$ for some CITIF $k$. Now if $\emptyset \neq A \cap L  \ni q$, then $k(m) \in I^+(q) \subset L$ for some $m \in \R$, as $L$ is a future set. Therefore all CITIFs from $k(m)$ are non-dominant. But $k$ itself is dominant as it dominates $c$ (as $I^- (c) \subset I^- (k) = A$ and $ q \notin I^- (c) $, again because $L$ is a future set). \hfill \qed

\bigskip

\v The upper horizon is an obstruction for the synopticity of the entire space-time:

\begin{Theorem}
If $(M,g)$ has noncompact Cauchy surfaces and an upper-shielded horizon, it is not synoptic. 
\end{Theorem}

\v {\bf Proof.} If $(M,g)$ is synoptic then no point of it is upper-shielded, as by Theorem \ref{buckelkurve}, we know that  from every point there is a curve $ k $ with $I^- (k) \neq M$. Then we pick $q \in M \setminus I^- (k) $ and look for some point $r \in I^+ (p) \cap I^- (q) $. Any future timelike curve joining $q $ with $r$ dominates $k$. \hfill \qed

\bigskip

\v Be aware of the fact that the reverse direction of the previous theorem does not hold: The de Sitter spacetime is not synoptic, but the future boundary is spacelike.

\bigskip

Finally, one could come up with a particularly easy horizon definition by spatial compactness:

\begin{Definition} 
In a globally hyperbolic manifold. A subset $A$ is called spatially (pre-)compact iff $A \cap S$ is (pre-)compact for every Cauchy surface $S$. A CITIF $c$ is called {\bf compact} iff $I^-(c)$ is spatially precompact. The compactness subset $C_M$ of $M$ is defined as the subset of all points $p \in M$ such that all CITIFs from $p$ are compact, and we define a compactness horizon $c_M$ as before as the boundary of a connected component of $CM$.
\end{Definition}

Of course this definition is only interesting for globally hyperbolic manifolds with noncompact Cauchy surfaces. Now there is a relation between the compactness horizon and the event horizon:

\begin{Theorem}[Hierarchy L-U-C-E of horizons in globally hyperbolic manifolds] 
In every globally hyperbolic manifold $(M, g)$, every non-dominating CITIF is compact, and the past of every compact CITIF contains a horizontal CITIF. Thus we have the inclusions $L_M \subset U_M , C_M$ and  $ U_M, C_M \subset E_M$.
\end{Theorem}

\v {\bf Remark.} It is an easy exercise to find counterexamples to any other inclusion between $C_M$ and any of $L_M$, $U_M$ and $E_M$, except for $U_M \subset C_M$. 

\v {\bf Proof.} After Theorem \ref{lues}, it remains to show $L_M \subset C_M \subset E_M$. To show the first, let $c: [0, \infty) \rightarrow M$ be a noncompact CITIF. We shall show that $c$ is dominating. Let $S$ be a Cauchy surface of $M$ containing $p:= c(0)$. As $I^-(c)$ is not spatially precompact, there is a sequence $p_n$ in $I^-(c) \cap S$ not converging in $S$. Choose a sequence $t_n$ such that $c(t_n) \in I^+(p_n)$ and choose timelike future curves $\gamma_n$ from $p_n$ to $c(t_n)$. We define $q:= c(1)$ and choose $r_n \in (\g_n([0, \infty)) \setminus J^+(q) ) \cap I^+(p)$ as well as future timelike curves $\rho_n$ from $p$ to $r_n$. Then we use the limit curve theorem in the globally hyperbolic manifold $K:= I^-(c) \setminus J^+(q)$ to show the existence of a limit curve $\rho$, which is not only a CITIF in $K$ but also in $M$, as the past of any possible endpoint $x$ would contain infinitely many of the $p_n$ and thus its intersection with $S$ would not be precompact. Now, $\rho$ as a curve in $I^-(c) \setminus J^+(q)$ is dominated by $c$.      

\v For the second inclusion, let $c$ be a compact CITIF, then either $c$ itself is horizontal, in which case the proof is complete. Or there is a sequence $c_n$ of CITIFs with $c_n \nearrow c$. We define $A := \bigcap_{k \ {\rm CITIF}, k<c} I^-(k)$. Note that the familiy of indices in this intersection is nonempty. this is a past set, and it is indecomposable because of minimality. Moreover, it is nonempty, as by Zorn's lemma we can write it as the intersection of a nonempty  monotonous chain $C_n$, and by compactness, for every Cauchy surface $S$, we get $S \cap C_n$ a monotonous chain of nonempty compacta, whose intersection is nonempty. So by the theorem of Kronheimer, Geroch and Penrose \cite{KGP}, $A$ is the past of a CITIF, which is necessarily horizontal. \hfill \qed

\bigskip

\v Secondly, let us now introduce a genuinely geometric, that is, not conformally invariant, notion with the same purpose of definition of future null infinity by defining {\em black holes}. As there is no precise and commonly accepted definition of this term except for the asymptotically flat case, we want to take a naive approach in which a black hole is characterized by two prominent features: it is {\em black and dangerous}. The first property is easily formalized as the requirement of being a future set, the second as the requirement of finite lifetime for all timelike future curves in the corresponding region.

\begin{Definition}
\label{black-hole}
Let $(M,g)$ be time-oriented Lorentzian. The {\bf black hole} of $M$ is the subset $BH_M$ of points $p$ in $M$ such that every future timelike curve through $p$ has finite length. \footnote{alternatively, one should think about variants of this requirement: e.g., that every causal geodesic has finite lifetime, every future b.a. curve has finite length, or requring that there is a uniform bound on these lengths etcetera. One could also define {\em a} black hole as a connected component of $BH_M$.}. The black hole $BH_M$ is called {\bf strong} iff for each point $p \in M \setminus BH_M$ there is a timelike geodesic\footnote{it could make sense to consider b.a.-curves instead of geodesics here.} of infinite length from $p$.   
\end{Definition}

\v Be aware of the fact that a black hole can have several connected components. Right from the definition, every black hole is a future set. Another advantage of this definition is the connection to incomplete geodesics which appear in the notion of singularities. If a manifold has a temporal function with the property M in the terminology of \cite{oM} then it cannot contain a black hole, as the integral curves of ${\rm grad} (t)$ are of infinite length. If a manifold of the form $ (N :=  I \times N , g:= - \l^2 dt^2 +g_t )$ with $(S_0,g_0)$ complete Riemannian, $\dot{g}_t/g_t$ globally bounded and $\l $ bounded from $0$ and from $\infty$ contains a black hole, it is not synoptic, for in this case $I$ has to be an interval bounded from above, so then $pr_{S_0} (I^+ (p))$ can be estimated uniformly in terms of $B_R(p)$, and one only has to choose two points in $S_0$ which are sufficiently far apart. Note that this latter case is not much more special than the spatially asymptotically flat case containing a black hole, as then all level surfaces of the Cauchy time function are complete, and the bounds hold for every level surface.

\v We want to establish connections between the causal notion of 'event horizon' and the geometric notion of 'black hole'. The definition corresponds to the exterior being the past of all timelike curves of infinite length. Those CITIFS could be defined as {\em geometric future null infinity}. It is an interesting question whether those pasts coincide with the pasts of complete {\em lightlike geodesics} as, traditionally, future null infinity is defined in terms of the latter. Obviously, not every event horizon bounds a black hole, as event horizons are conformally invariant and every globally hyperbolic manifold is conformally invariant to a future causally complete one (cf. \cite{hS}, p. 258). More exactly we have

\begin{Theorem}[compare with \cite{hS}] 
\label{conformal}
\begin{enumerate}
\item{Every globally hyperbolic manifold is conformally equivalent to another one which is b.a.-complete and null geodesically complete.}
\item{For every $E>0$, every globally hyperbolic manifold is conformally equivalent to another in which all nonspacelike curves have length smaller than $E$.}
\end{enumerate}
\end{Theorem}

\v {\bf Proof.} We include a proof of the statement as the proof appearing in the article of Seifert uses tacitly the following lemma guaranteeing the existence of useful time functions:

\begin{Lemma} 
Let $S$ be a Cauchy surface in a globally hyperbolic manifold $M$ and let $f \in C^0 (S) $. Then there is a smooth Cauchy time function $t$ on $M$ with $t(x) > f(x)$ for all $x \in S$. In particular, as $f$ can be chosen to be proper, there is a smooth Cauchy time function $t$  on $M$ with $ S \cap t^{-1} ((- \infty,D))  $ compact for all $D \in \R$.  
\end{Lemma}

\v {\bf Proof of the lemma.} Let $t_0 $ be a smooth Cauchy time function such that $S= t_0^{-1} (\{ 0 \} )$. Let the compact subsets $U_n \subset U_{n+1}$ cover S. We want to reach 

\bea
\label{more}
t\vert_{U_n} \geq n.
\eea

\v We cover each $U_{n+1} \setminus U_n$ locally finitely by sets $A_i^n := I^+ (p_i^n)$ such that $A_i^n \cap U_{n-1} = \emptyset$ and $A_i^n \cap (U_{n+3} \setminus U_{n+2}) = \emptyset$. Then let $t_i^n$ be a smooth Cauchy time function on $A_i^n$ and $\tau_i^n := exp (t_i^n)$ on $A_i^n$ and zero on the complement of $A_i^n$. Then $t^n:= \sum a_i^n \cdot \tau_i^n \in C^{\infty} (M, \R)$ is a time function on an open subset containing $I^+ (U_{n+1} \setminus U_n)$ and, for an appropriate choice of $a_i^n$, we have $t^n \vert_{U_{n+1} \setminus U_n} \geq n$. Finally, $t:= t_0 + \sum_{i=1}^{\infty}$ is well-defined and smooth on $M$ by standard arguments, it is a Cauchy time function as a sum of a Cauchy time function and a time function (recall that the latter is defined by monotonicity along future causal curves and the former additionally by surjectivity along $C^0$-inextendable future causal curves), and it satisfies Eq. \ref{more}, proving the statement. 

\bigskip

\v {\bf Proof of the Theorem.} Let $S$ be a Cauchy surface in $M$ and $t$ a smooth Cauchy time function as in the lemma. $t$ induces a splitting $ug = - a dt^2 - ag_t^2$ of any conformal multiple $ug$ of $g$. We get as geodesic equation in $(M,ug)$ for the $t$ coordinate along a geodesic curve $c$:

$$  \frac{d^2 t}{ds^2}  = \frac{a_{,m}}{a} \frac{dx^m}{ds} \frac{dt}{ds} - \frac{\dot{a}}{2a}  (\frac{dt}{ds} )^2    - \frac{1}{2} (\dot{g}_{mn} + \frac{\dot{a} }{a} g_{mn} )   \frac{dx^m}{ds}   \frac{dx^n}{ds},$$

 \v where $s$ is the affine parameter. With the ansatz $a = a(t)$ the first term vanishes. If we consider a b.a. curve we get a bounded real function $D$ as an additional aditive term on the right hand side. For the geodesic case just replace $D$ by $0$ from now on. Because of compactness of the $t$ level sets intersected with $J^+(S)$ there is an $f(t) $ with $(\dot{g}_{mn} + f(t) g_{mn} ) x^m x^n \geq 0$, that is to say, $\dot{g} + f(t) \cdot g $ positive definite. Now if $\frac{d (log (a(t)))}{dt} \geq max \{  0, f(t)  \}    $ then $\frac{d^2 t}{ds^2} \leq D$, but as $ \frac{ dt}{ds} >0$ we get $s \geq (\frac{dt}{ds})^{-1} \vert_{s=0} \cdot t$ (geodesics) or $s \geq D (\frac{dt}{ds})^{-1} \vert_{s=0} \cdot t^2  $ (b.a. curves), so in either case, for bounded $s$, $t$ is bounded as well, thus $(I^+ (S), a \cdot g )$ is future geodesically complete and b.a.-complete. Now choose another Cauchy hypersurface $S_0 \subset I^- (S) $ and perform the same procedure in the past direction constructing a conformal factor $a_0$ on $I^- (S_0)$, and finally interpolate in any manner between $a_0$ and $a$, showing the first assertion of the theorem. The second one is easier, if one uses the previous lemma one has only to choose the factor in the compact balls so that the maximum over the resulting factors before $-dt^2$ is integrable, or equivalently the lightlike affine parameter (using the compactness of lightlike directions additionally). \hfill \qed

\bigskip

\v The question is, does the presence of an event horizon {\em in a Ricci flat manifold} imply the presence of a black hole? Again the answer is no, as may be seen by considering the flat hexagon which is the Penrose diagram of Kruskal spacetime: It is even flat, but it is a black hole in itself, and the event horizon does not bound a black hole. One can modify this to get an example of a {\em maximal} Lorentzian manifold with the same sort of behaviour by considering a smooth function $f$ on the hexagon diverging to infinity towards its boundary and then choosing $2 + {\rm sin} \o f $ as a conformal factor. Of course, the Ricci-flatness is lost by this procedure. So the new question could be: {\em Can there be an eager-beaver censor in a Ricci-flat time-oriented maximal Lorentzian manifold in the sense that is geodesically complete and still contains a horizon of some kind?} And conversely, the question could be whether a maximally globally hyperbolic Ricci-flat manifold which is geodesically incomplete contains a horizon of some sort. The answer to the last question is 'no', as Taub-NUT space-time provides a counterexample - the pasts of all of its CITIFs coincide.  

\bigskip

\v Finally, we introduce another, possibly useful, notion of black hole horizon inspired by Penrose's 1978 paper \cite{P}. His condition CC4 --- 'no $\infty$ - TIP contains a singular TIP' --- would imply $BH_M \subset V^-_M $ in the terminology below. We define the {\bf future visibility subset} 

$ V_M^+ := \{ p \in M \vert \forall c {\rm \ CITIF \  from \ } $p$, \forall h {\rm \ curve \ of \ infinite \ length : \  } I^- (c) \not \supset h \}  $

\v as well as the {\bf past visibility subset}

$ V_M^- := \{ p \in M \vert \forall c  {\rm \ CITIF \  from \ } $p$, \forall h {\rm \ curve \ of \ infinite \ length : \  } I^- (h) \not \supset c \}  $

\v we have always $V_M^+ \subset BH_M$, right from the definition. We show:

\begin{Theorem}

If $(M,g)$ is globally hyperbolic and if, for each $x \in M$, $I^+ (x)$ admits a real $C^1$ function $f$ such that there are $D,G >0$ with $ - G^2 < \langle {\rm grad} f (p) , {\rm grad} f (p) \rangle < - D^2 $ for all $p \in I^+ (x) $, \footnote{Compare with Theorem 3 of \cite{oM} where for each Cauchy surface $S$, a time function $t$ on $I^+ (S)$ is constructed with $\langle {\rm grad} t (p) , {\rm grad} t  (p) \rangle $ bounded on each of the levelsets of $t$. In a forthcoming paper \cite{o2}, the condition of this theorem will be shown to be valid in an arbitrary globally hyperbolic manifold.} then $V_M^+ = BH_M $.

\end{Theorem}

%\v {\bf Remark:} Compare with the situation in 2-dimensional Kruskal spacetime: Here, the Killing vector field is $X:= v \partial_v - u \partial_u $ in the standard null coordinates. It is not integrable itself but pointwise proportional to ${\rm grad} (r)$. Still, the bounds required in the theorem do not hold.

\bigskip

\v {\bf Proof.} Let $ p \in BH_M $, let $k$ be the maximal integral curve of $ {\rm grad} f $ from $p$, this is a CITIF parametrized on $[0, S) $ with $S \leq \infty$. Then $(f \o k)' = \langle \dot{k}, {\rm grad} f  \rangle =  - \langle {\rm grad} f , {\rm grad} f \rangle =   - \langle  \dot{k} , \dot{k} \rangle $ and therefore 

$\int_0^S \vert (f \o k)' (s)  \vert ds \leq G \cdot \int_0^{S} \sqrt{ \vert (f \o k)' (s)  \vert } ds =   G \cdot \int_0^{S} \sqrt{- \langle \dot{k} (s), \dot{k} (s) \rangle } ds $

\v therefore $ f \vert_{I^+(p)}$ is bounded by a real number $E$. Now assume that there is a future timelike curve $h$ of infinite length in $ I^- (c) $, parametrized by arc length for simplicity. Then, as $h(n)$ is in the past of the image of $c$ for all $n$, we can find a timelike geodesic curve $k_n: [0, 1] $ of length $> n$  from $h(0)$ to some point of the image of $c$. We write $X= {\rm grad} f$ for short. Then we get for the length $l(k_n)$ of $k_n$:

\bean
  l(k_n) &=&  \int_0^1 \sqrt{\langle \dot{k}_n (s) , \dot{k}_n (s) \rangle } ds \leq \int_0^1 \frac{\langle \dot{k}_n (s), X  (k_n (s)) \rangle}{-\langle X (k_n (s) ) , X (k_n (s)) \rangle} ds \\  &\leq& D \int_0^1 \langle \dot{k}_n (s), X  (k_n (s)) \rangle ds  .
    \eean

\v Here the first inequality is due to the inverse Cauchy-Schwarz inequality. Consequently, we have, for all $n \in \N$,

\bean
l(k_n) \leq - D \int_0^1 \langle {\rm grad} f (k_n(s)), \dot{k}_n (s) \rangle dt \leq f(k_n(1)) - f(h(0)) \leq E - f(h(0)) 
\eean

\v which is a contradiction \hfill \qed

\bigskip

\v Finally, let us compare the previous notions with the ones using a Killing vector field. Recall that a (local) Killing horizon is a null hypersurface $N$ whose tangent bundle contains $X(N)$ for some (local) Killing vector field $X$ (around $N$). We will consider a narrower notion.

\begin{Definition}
A subset $A \subset M$ of a globally hyperbolic manifold is {\bf spatially bounded} if there is a positive number $R$ such that, for every smooth Cauchy hypersurface $C$, the intersection of $A$ with $C$ has diameter less or equal to $R$ in the Riemannian metric of $C$. A subset $A \subset M$ of a globally hyperbolic manifold is {\bf spatially compact} if, for some (hence any) Cauchy time function $t$, the intersection of $A$ with each level set of $t$ is compact. A {\bf strong Killing horizon} is a hypersurface $S$ in $M$ which bounds a connected, spatially bounded future set on which a Killing field $X$ is spacelike and also bounds a subset on which $X$ is timelike. An {\bf ultrastrong Killing horizon} is a hypersurface $S$ in $M$ which bounds a connected, spatially bounded and spatially compact future set on which a Killing field $X$ is spacelike and also bounds a subset on which $X$ is timelike. 
\end{Definition}

%The condition to be a null hypersurface is necessary as e.g. in Minkowski space otherwise any timelike cylinder would be a Killing horizon by considering spatial rotations around an axis composed with timelike translations along this axis. Also the other restrictive conditions are necessary considering the following examples. 

\v Obviously, in this terminology, every strong Killing horizon is the boundary of a spatially bounded set, and every ultrastrong Killing horizon is the boundary of a spatially compact and spatially bounded set. The Schwarzschild horizon is a strong Killing horizon, but it is not ultrastrong. We would like to show that every strong Killing horizon is an event horizon. Unfortunately, that is not true. A counterexample is easily found on $\R^2$ by the usual technique of cone ballet on the $x$ axis: Consider coordinates $(t,x)$ and $u:= x+y, v:= x-y$ and define the metric $g := \phi (x) \cdot dv (du + f(x) dv )$, for some smooth monotonously non-increasing surjective $f: \R \rightarrow [0,2) $. This metric is globally hyperbolic, as its causal cones are narrower than the corresponding causal cones of Minkowski spacetime. Moreover, $X := \partial_t$ is a Killing vector field which is spacelike on the left half $L:= f^{-1} ((1,2))$ of $\R^2$ which is a future set. And by an appropriate choice of the conformal factor $\phi$, $L$ can be made spatially bounded. But there is no shielded or event horizon in $(\R^2, g)$, as the past of every future timelike curve is properly contained in the past of the same curve composed with the flow of $X$. Moreover, the manifold is synoptic. So {\em the existence of a strong Killing horizon does not imply the existence of a horizon of any kind}. Still the following question remains unanswered: {\em Does the existence of an ultrastrong Killing horizon in a Ricci-flat maximal Lorentzian manifold $(M,g)$ imply the existence of an event horizon in $M$?} In any case, given the nongenericity of Killing horizons, it does not seem very likely to the author that the notion of Killing horizons can contribute much to the questions around the weak cosmological censorship conjecture in their above formulation.

% Equally, there are compact horizons which are not event horizons: Consider the continuous even function $f: \R \rightarrow [0,2]$ with $f(x) := 0$ for all $\vert x \vert \geq 3$, $f(x) = x+3$ for $x \in [-3,-2]$, $f(x) := 1$ for $x \in [-2,-1]$ and $f(x) := x-2$ for $x \in [-1,0]$. Then the region $ A:= \{ p=(t,x) \in \R^{1,1}  \vert  0<t< f(x) $ is globally hyperbolic as a causal subset of Minkowski spacetime. It contains a compact horizon (given by the intersections of the lines $\{ t= -2 \pm x \}$ with $A$) which is not an event horizon. By modifying the previous example by a conformal factor as in Theorem \ref{conformal} below, one even obtains an example of a {\em b-complete globally hyperbolic manifold containing a compact horizon which is not an event horizon.}

\end{document}